\documentclass[twocolumn,prb,showpacs,amsmath,amssymb]{revtex4}
\usepackage{graphicx}
\usepackage{dcolumn} 
\usepackage{bm}      
\usepackage{mathrsfs}
\usepackage{ulem}
\usepackage{fdsymbol}

\usepackage{fancyhdr}
\usepackage{color}
\usepackage[usenames,dvipsnames]{xcolor}
\newcommand{\red}{\color{Red}}
\newcommand{\green}{\color{Green}}

\newcommand{\magenta}{\color{Magenta}}

\begin{document}

\title{Thermodynamics of a spin-1/2 XYZ Heisenberg chain with a Dzyaloshinskii-Moriya interaction}

\author{Bin Xi$^{1,2}$}
\altaffiliation{These two authors contributed equally to this work.}
\author{Shijie Hu$^3$}
\altaffiliation{These two authors contributed equally to this work.}
\author{Qiang Luo$^{2}$}
\author{Jize Zhao$^{4}$}\email{jizezhao@gmail.com}
\author{Xiaoqun Wang$^{5,6,2}$}\email{xiaoqunwang@ruc.edu.cn}
\affiliation{$^1$College of Physics Science and Technology, Yangzhou University, Yangzhou 225002, China}
\affiliation{$^2$Department of Physics and Beijing Laboratory of Opto-electronic Functional Materials $\&$ Micro-nano Devices, Renmin University of China, Beijing 100872, China}
\affiliation{$^3$Department of Physics and Research Center Optimas, Technical University Kaiserslautern, 67663 Kaiserslautern, Germany}
\affiliation{$^4$Institute of Applied Physics and Computational Mathematics, Beijing 100088, China}
\affiliation{$^5$Department of Physics and Astronomy, Shanghai Jiao Tong University, Shanghai 200240, China}
\affiliation{$^6$China and Collaborative Innovation Center for Advanced Microstructures, Nanjing 210093, China}

\date{\today}

\begin{abstract}
We study the thermodynamics of a spin-1/2 XYZ Heisenberg chain with a Dzyaloshinskii-Moriya interaction. 
This model describes the low-energy behaviors of a one-dimensional two-component bosonic model 
with a synthetic spin-orbit coupling in the deep insulating region. In the limit $U^\prime/U\rightarrow\infty$,
where $U$ is the strength of the onsite intracomponent repulsion and $U^\prime$ is the intercomponent one,
we solve our model exactly by Jordan-Wigner transformation, and thus provide a benchmark for our following numerical approach.
In other cases, we calculate the entropy and the specific heat numerically by the transfer-matrix renormalization group method.
Their low-temperature behaviors depend crucially on the properties of the zero-temperature phases. 
A refined ground-state phase diagram is then deduced from their low-temperature behaviors.
Our findings offer an alternative way to detect those distinguishable phases experimentally.

\end{abstract}
\maketitle

\section{Introduction}
One-dimensional (1D) quantum magnetism remains an active research area in condensed matter physics because of their intriguing properties
arising from strong quantum fluctuations\cite{Giamarchi}.
In this area, the 1D spin-$1/2$ antiferromagnetic (AF) Heisenberg chain is a prototypical model, the ground state of which  
is a Tomonaga-Luttinger liquid (TLL) \cite{Haldane80}. It has gapless elementary excitations and is relevant to a variety  of quasi-1D
magnetic materials\cite{Motoyama96,Hirakawa70,Kono15}. However,  its properties may change significantly in the presence
of anisotropy\cite{BAXTER1, ERCOLESSI1, CAO1}.

In addition to abundant quasi-1D  materials, ultracold atomic systems in optical lattices
have already become an important platform to simulate quantum spin systems.
Spin-spin interaction using controlled collisions was first proposed\cite{Jaksch99} theoretically in 1999
and later successfully realized in experiments with $^{87}$Rb atoms\cite{Collins03}.
In these experiments, the two hyperfine states $|F = 1,m_F = -1\rangle$ and $|F = 2,m_F = -2\rangle$ of $^{87}$Rb atoms are treated
as up and down spins\cite{Collins03}, respectively. This two-component boson mixture soon attracted a great deal of interest.
Duan and coworkers suggested that the Hamiltonian of this two-component system can be mapped into a
spin-1/2 XXZ Heisenberg model\cite{Duan03}. Its ground state is ferromagnetic (FM) when the intercomponent repulsion $U'$ is
much larger than the intracomponent one $U$, while it is AF when $U'\ll{U}$.
These studies have provided us valuable information to understand some long-standing problems in condensed matter physics.
After these pioneering works, more complicated spin models have been proposed in the context of optical lattices. For example,
it was demonstrated that XYZ Heisenberg models can be implemented with p-orbit bosons\cite{PINHEIRO1} in one dimension, and
with Rydberg atoms in two dimensions\cite{GLAETZLE1,BIJNEN1}.

Recently, a synthetic spin-orbit coupling (SOC), or equivalently, gauge field, was successfully realized in experiments and a variety of
phases as well as phase transitions were observed\cite{Lin11,Wang12,Cheuk12,Zhang12}. 
These experiments have spurred great interest in studying the artificial SOC as well as gauge field 
in ultracold systems\cite{Radic12, Cai12, Cole12, Gong15, Zhou13, Fu14,Zhao14A, Zhao14B, Piraud14, Xu14, Peotta14, 
Hamner15, Zhao15, Syzranov14, Orignac16, Wu16}. 
In the deep insulating region,
such an SOC can be approximated \cite{Cole12, Zhao14B} by the Dzyaloshinskii-Moriya (DM) interaction\cite{Dzyaloshinsky58,Moriya60}.
In many magnetic materials, DM interaction plays a key role in understanding a variety of exotic magnetic features,
e.g. spiral magnetism\cite{Kimura07, Radic12, Cole12, Cai12}, skyrmion\cite{Mhlbauer09,Yu10,Heinze11,Nagaosa13,Fert13}.
Therefore, it is expectable that rich magnetic structure can be experimentally observed in ultracold 
atomic systems with the SOC.

The SOC realized in 2011 has equal weight of Rashba and Dresselhaus terms\cite{Lin11}. Thus it is along one direction in real space.
Loaded into 1D optical lattice\cite{Hamner15}, the low-energy dynamics of such spin-orbit-coupled bosons 
can be modeled by the Hamiltonian\cite{Zhao14A}
\begin{eqnarray}
\hat{\mathcal{H}}_{\textrm{boson}} & = & \hat{\mathcal{K}} + \hat{\mathcal{T}}_{\text{soc}} + \frac{U}{2} \sum_{i\tau} \hat{n}_{i\tau} ( \hat{n}_{i\tau} - 1) \nonumber \\
&   & + U^{\prime} \sum_i \hat{n}_{i\uparrow} \hat{n}_{i\downarrow},
\label{OriginH}
\end{eqnarray}
where $\hat{\mathcal{K}} = -t\sum_{i\tau} (\hat{c}^{\dagger}_{i\tau} \hat{c}_{i+1\tau} + \textrm{H.c.})$ is the hopping term
between the nearest-neighbor sites with the hopping integral $t$.
$\hat{\mathcal{T}}_{\text{soc}} = -\lambda\sum_i (\hat{c}^{\dagger}_{i\uparrow} \hat{c}_{i+1\downarrow} - \hat{c}^{\dagger}_{i\downarrow} \hat{c}_{i+1\uparrow} + \textrm{H.c.})$ is the SOC. The strength of the SOC $\lambda$ can be controlled by the laser frequency.
$\hat{c}^\dagger_{i\tau}$ ($\hat{c}_{i\tau}$) is the creation (annihilation) operator of bosons at
site $i$ with spin $\tau$. $\tau$ takes $\uparrow$ and $\downarrow$, representing two internal states of atoms.
$U$ is on-site intracomponent repulsion and $U'$ is the intercomponent one.
$\hat{n}_{i\tau}=\hat{c}^\dagger_{i\tau}\hat{c}_{i\tau}$ is the boson number operator with spin $\tau$ at
site $i$. $\mu$ is the chemical potential to control the filling factor. At unit filling and in strong coupling limit $t,\lambda \ll U, U^\prime$,
this model can be effectively written as a spin-$\frac{1}{2}$ XYZ Heisenberg chain with a DM interaction (see Ref. \cite{Zhao14B} for more details).
By setting $t=J\cos{\theta}$, $\lambda=J\sin{\theta}$,  it reads:
\begin{eqnarray}
\mathcal{\hat H} & = & \frac{4J^2}{U}\left[ \left(-2+\frac{U}{U^{\prime}}\right)\cos 2\theta\sum_i {\hat S}^z_i {\hat S}^z_{i+1}\right. \nonumber \\
&   & - \frac{U}{U^{\prime}}\cos 2\theta \sum_i {\hat S}^x_i {\hat S}^x_{i+1}- \frac{U}{U^\prime}\sum_i {\hat S}^y_i {\hat S}^y_{i+1} \nonumber \\
&   & - \left. \sin2\theta \sum_i ({\hat S}_i^z {\hat S}_{i+1}^x - {\hat S}_i^x {\hat S}_{i+1}^z)\right],
\label{HEFF}
\end{eqnarray}
where $\hat{S}_i^{\nu} = \sum_{\tau\tau^{\prime}} \hat{c}^{\dagger}_{i\tau} \hat{\sigma}_{\tau\tau^{\prime}}^{\nu} \hat{c}_{i\tau^\prime}$
are the pseudo-spin operators with $\hat{\sigma}^{\nu}$ Pauli matrix and $\nu=x,y,z$.

The Hamiltonian (\ref{HEFF}), or equivalently Hamiltonian (\ref{OriginH}) at unit filling in the strong coupling limit, has been studied by 
several groups using density-matrix renormalization group (DMRG) method
in combination with some analytic methods\cite{Zhao14A,Zhao14B,Piraud14,Xu14,Peotta14}.
For $U^\prime=U$, the DM interaction can be eliminated by a site-dependent rotation of the spin operators, resulting in an isotropic
Heisenberg chain with FM coupling\cite{Cai12}. In this sense, the SOC becomes trivial in such a case.
However, when $U^\prime\neq U$, the DM interaction cannot be simply eliminated\cite{Zhao14A} and several phases have been predicted.
For $U^\prime>U$, there are a gapped FM phase, a gapped AF phase,
and in between a TLL phase with a chiral order\cite{Zhao14B, Piraud14, Xu14}(without ambiguity, we will call it TLL phase below). 
The transition from the FM (AF) phase to the TLL phase is of first order\cite{Piraud14,Peotta14}.
For $U^\prime<U$, a gapless paramagnetic phase and a gapful FM phase are found\cite{Zhao14A, Piraud14, Xu14}.
The transition between these two phases is of Berezinskii-Kosterlitz-Thouless(BKT)\cite{KBT} type. However, due to the limit of
numerical accuracy and finite-size effect, the critical point has not been determined accurately so far.

The abovementioned studies are all limited to zero temperature.
The properties of the Hamiltonian (\ref{HEFF}) at finite temperature remain unknown yet.
In particular, when approaching zero temperature, what are the asymptotic behaviors of some typical
quantities such as the entropy and the specific heat?
Understanding these questions is remarkably important for determining the phase diagram experimentally.
On the other hand, Hamiltonian (\ref{HEFF}) is quite general although it originates
from the context of ultracold systems. We believe that it is qualitatively related to some
quasi-1D  materials, such as Copper benzoate\cite{DENDER1}, Cs$_2$CoCl$_3$\cite{Kenzelmann1},
$\text{CuCl}_2 \cdot$2(dimethylsulfoxide) (CDC) \cite{Kenzelmann04,Chen07}, copper pyrimidine \cite{Feyerherm00,Zvyagin04,Zvyagin05} and $\text{Yb}_4\text{As}_3$ \cite{Kohgi01}.

In this work, we study the thermodynamics of the Hamiltonian (\ref{HEFF})
with the transfer-matrix renormalization group (TMRG) method \cite{Wang97}.
TMRG is a powerful numerical method for studying the thermodynamics of 1D quantum systems. 
It treats infinitely large systems directly, and thus there is no finite-size effect. We refer the reader
to references \cite{Bursill96,Wang97,Wang00,Xiang98} for more details. During the TMRG iterations,
$1000\sim{2000}$ states are kept in most cases. The truncation error is less than $10^{-12}$ in all calculations.
Particularly, we use an additional reorthogonalisation procedure after the left and right eigenvectors of the reduced density matrix are
obtained. This allows us to keep more states and thus improve accuracy \cite{Honecker11}.
In Hamiltonian (\ref{HEFF}), the particle fluctuation is completely suppressed.  
Therefore, if we focus only on the magnetism in spin-orbit-coupled bosonic systems, 
Hamiltonian (\ref{HEFF}) is a more appropriate model for numerical simulations than Hamiltonian (\ref{OriginH}).
For simplicity, we set $4J^2/U$ as the energy unit. One can immediately see
that Hamiltonian (\ref{HEFF}) has a period of $\pi/2$ in $\theta$ by performing 
the transformation  $\hat{S}_{2i+1}^x\rightarrow-\hat{S}_{2i+1}^x$, $\hat{S}_{2i+1}^y\rightarrow \hat{S}_{2i+1}^y$ and $\hat{S}_{2i+1}^z\rightarrow-\hat{S}_{2i+1}^z$.
Moreover, one can interchange $t$ and $\lambda$ in Hamiltonian (\ref{OriginH})\cite{Zhao14A, Zhao15}, so we only need to consider the parameter region
$\theta\in[0, \pi/4]$ since the properties in the region $(\pi/4, \pi/2]$ are readily available. It is straightforward to verify this 
in Hamiltonian (\ref{HEFF}) by using the fact $\sin{\theta}=\cos(\frac{\pi}{2}-\theta)$.

The paper is organized as follows: in Section II, we begin our study in the exactly solvable limit $U^\prime/U \rightarrow \infty$.
In Section III, we consider the region $U^\prime/U>1$. The phase transition points are obtained through the isentropic map.
In the low-temperature limit, the asymptotic behaviors of the specific heat and the entropy in different phases are compared.
In Section IV, we focus on the region $U^\prime<U$. We determine the critical point from the entropy.
In Section IV, we give our conclusions.

\begin{figure}
\centering
\includegraphics[width=.9\linewidth]{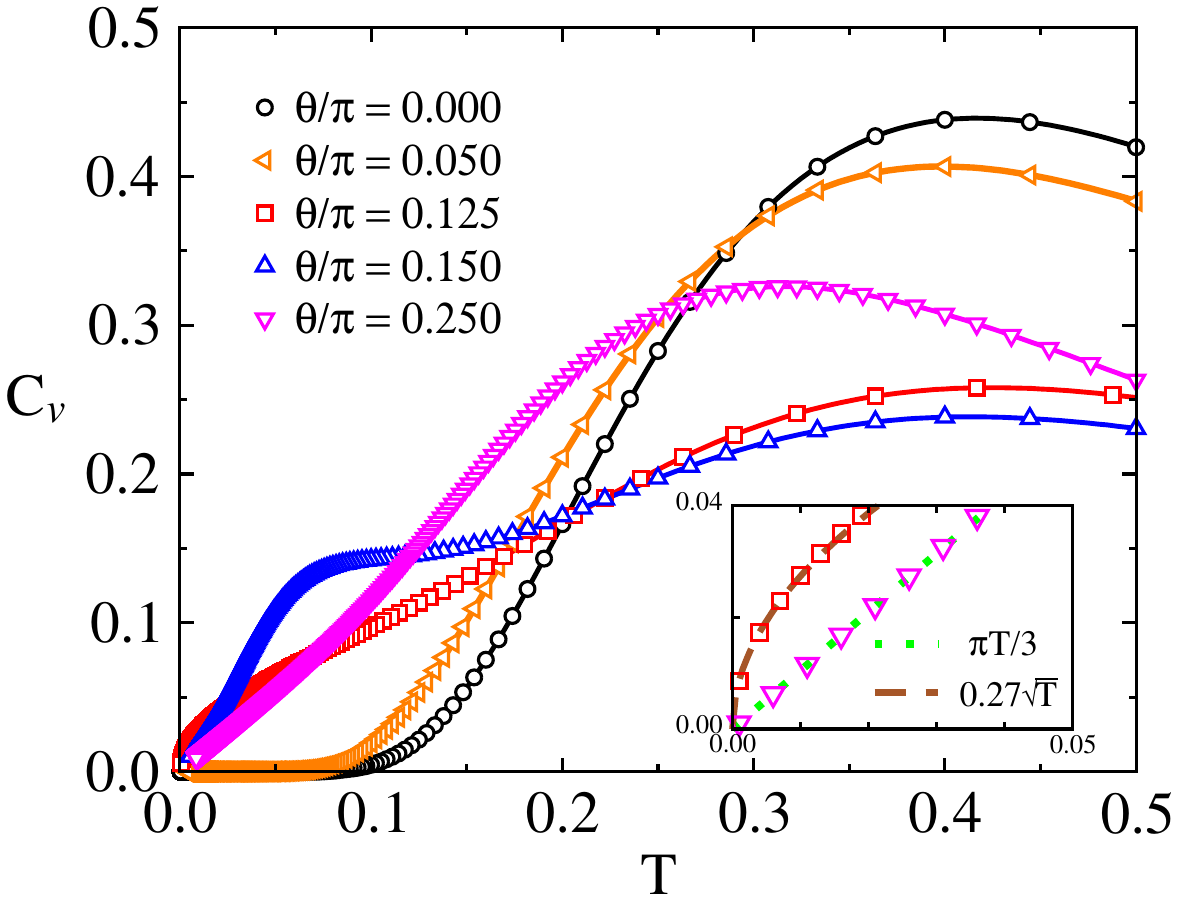}
\caption{(color online). Exact solutions (lines) and TMRG results (symbols) of the specific heat $C_\nu$ are plotted
as a function of temperature $T$ for a variety of $\theta$.
Inset shows asymptotic behavior of the specific heat at low temperature for two different cases:
1) $\theta=\pi/4$ ({\magenta $\triangledown$}) which is in the TLL phase, the specific heat is proportional to $T$ and
2)  $\theta=\pi/8$ ({\red $\square$}) which is the transition point from the TLL phase to the gapped FM phase,
the specific heat behaves as  $\sqrt{T}$.}
\label{fig1}
\end{figure}

\section{Exactly Solvable Case}
In the limit $U'/U\rightarrow \infty$, Hamiltonian (\ref{HEFF}) is reduced to
\begin{eqnarray}
\mathcal{\hat{H}} & = & -2\cos 2\theta \sum_i \hat{S}^z_i \hat{S}^z_{i+1} \nonumber \\
                     &   & -\sin2\theta\sum_i (\hat{S}_i^z \hat{S}_{i+1}^x - \hat{S}_i^x \hat{S}_{i+1}^z).
\label{eslH}
\end{eqnarray}
One can immediately see that at $\theta = 0$ the Hamiltonian (\ref{eslH}) is just an Ising model with a FM ground state,
while at $\theta=\pi/4$ it is equivalent to an isotropic XY model, which has a TLL ground state. For general $\theta$,
the Hamiltonian (\ref{eslH}) can be transformed into a Kitaev chain by the Jordan-Wigner transformation\cite{DERZHKO}, which is exactly solvable,
leading to $\mathcal{\hat H}=\sum_k E_k(\hat{\Lambda}^{\dagger}_k \hat{\Lambda}_k -1/2)$ with the energy 
dispersion $E_k=\cos2\theta -\sin2\theta\sin k$ and $\hat{\Lambda}^{\dagger}_k$ ( $\hat{\Lambda}_k$) the creation (annihilation) operator 
of fermions with the momentum $k$ (see Appendix A for more details). One can notice that the system undergoes 
a quantum phase transition from a gapped phase into a gapless one at $\theta=\pi/8$.  
The thermodynamic properties of the Hamiltonian (\ref{eslH}) can then be exactly calculated from the partition
function $Z$ in a standard way. For example, the specific heat $C_\nu$ can be expressed as
\begin{eqnarray}
C_\nu &  = & \beta^2\frac{\partial ^2 \ln Z}{\partial \beta^2} \nonumber \\
    &  = & \displaystyle\frac{1}{2\pi}\int_{-\pi}^\pi dk (\beta E_k/2)^2 \cosh^{-2} (\beta E_k/2),
\end{eqnarray}
with the inverse temperature $\beta =1/T$. $C_\nu$ can be evaluated after a numerical integration.
The results are shown in Fig.~\ref{fig1} together with our TMRG results.
One can see that our TMRG results agree perfectly with the exact ones, verifying the precision of the TMRG data.

The low-temperature behavior of the specific heat reveals distinguishable features for different values of $\theta$.
At $\theta=0$, the system is just a classical Ising chain and $E_k\equiv 1$.
One can easily obtain $C_\nu=(\beta/2)^2 \cosh^{-2} (\beta/2)$, which can be approximate to $T^{-2}\exp(-1/T)$ under low-$T$ limit.
For $0<\theta < \pi/8$, though the ground state is also an Ising-type FM phase, the low-temperature behavior is different.
Here the low-energy excitations are the gapful magnons, whose dispersion can be approximately written as
\begin{eqnarray}
\epsilon_q=\Delta+\frac{\sin{2\theta}}{2} ~q^2+O(|q|^3),
\end{eqnarray}
where $q=k-\pi/2$ and $\Delta=\cos2\theta-\sin2\theta$ is the energy gap between the ground state and the first excitation at $k=\pi/2$.
The $q^2-$dependence of the magnon dispersion results in
$C_\nu\sim T^{-3/2}\exp(-\Delta/T)$ for $T\ll\Delta$\cite{Xiang98}.
These two different exponential behaviors are shown in Fig.~\ref{fig1} with $\theta=0$ and $0.05\pi$.
At $\theta=\pi/8$, a phase transition takes place between the gapful FM phase and the TLL.
At this point, the gap is closed, and the dispersion is proportional to $q^2$.
Therefore, one has the density of states $g(E_{\pi/2})\sim dk/dE|_{k=\pi/2}\sim 1/\sqrt{E_{\pi/2}}$.
It turns out that the free energy F reads
\begin{equation}
\displaystyle\int dE\frac{E g(E)}{\exp(E/T)+1} \sim T^{3/2},
\end{equation}
which leads to a $T^{1/2}$-dependence of the specific heat as shown in the inset of Fig.1.
In the TLL phase corresponding to $\theta>\pi/8$,  one has effectively a Fermi momentum  $k_F=\arcsin(\tan2\theta)$,
which shifts from $\pi/2$ towards $0$ with $\theta$ increasing further from the transition point.
As a consequence, the specific heat exhibits a bump  at low temperature and becomes linear in the very low $T$ regime.
An example is given for $\theta/\pi = 0.15$ in Fig.1. The bump reflects the contribution from the 
excitations with the dispersion deviating from the linearity and suggests a  crossover  
from an ideal TLL with linear excitations and  others with $k^2$-dependent excitations. 
The bump shifts to higher temperature as the $\theta$ increases and is eventually
absorbed by the peak of the specific heat at $\theta=\pi/4$. In the TLL phase,  the dispersion of the low-energy 
excitations is proportional to the momentum,
which results in a $T^2$-dependence of free energy at very low temperature so that one has $C_\nu/T=\pi/3v$ 
with $v$ the spin-wave velocity\cite{Sirker12}. In our model, $v=\sqrt{ - \cos 4 \theta}$. 
Therefore  the specific heat has the following low-temperature behavior,
\begin{eqnarray}
 C_\nu=\frac \pi {3\sqrt{ - \cos 4 \theta}}T.
 \label{CvT}
\end{eqnarray}
The inset of Fig. 1 illustrates this behavior for  $\theta=\pi/4$ as  compared  with TMRG results.

\section{$\mathbf{U^\prime/U > 1}$}

After benchmarking our TMRG method, we now turn to our main task, the thermodynamics in the anisotropic interacting case, i.e.,
$U^\prime/U$ is finite but $U^\prime/U \neq 1$. Under this condition, the Hamiltonian in general is not exactly solvable,
and thus we resort to the TMRG method to study it. In this section, we focus on $U^\prime/U>1$.
\subsection{Entropy}
\begin{figure} [ht!]
\centering
\includegraphics[width=.9\linewidth]{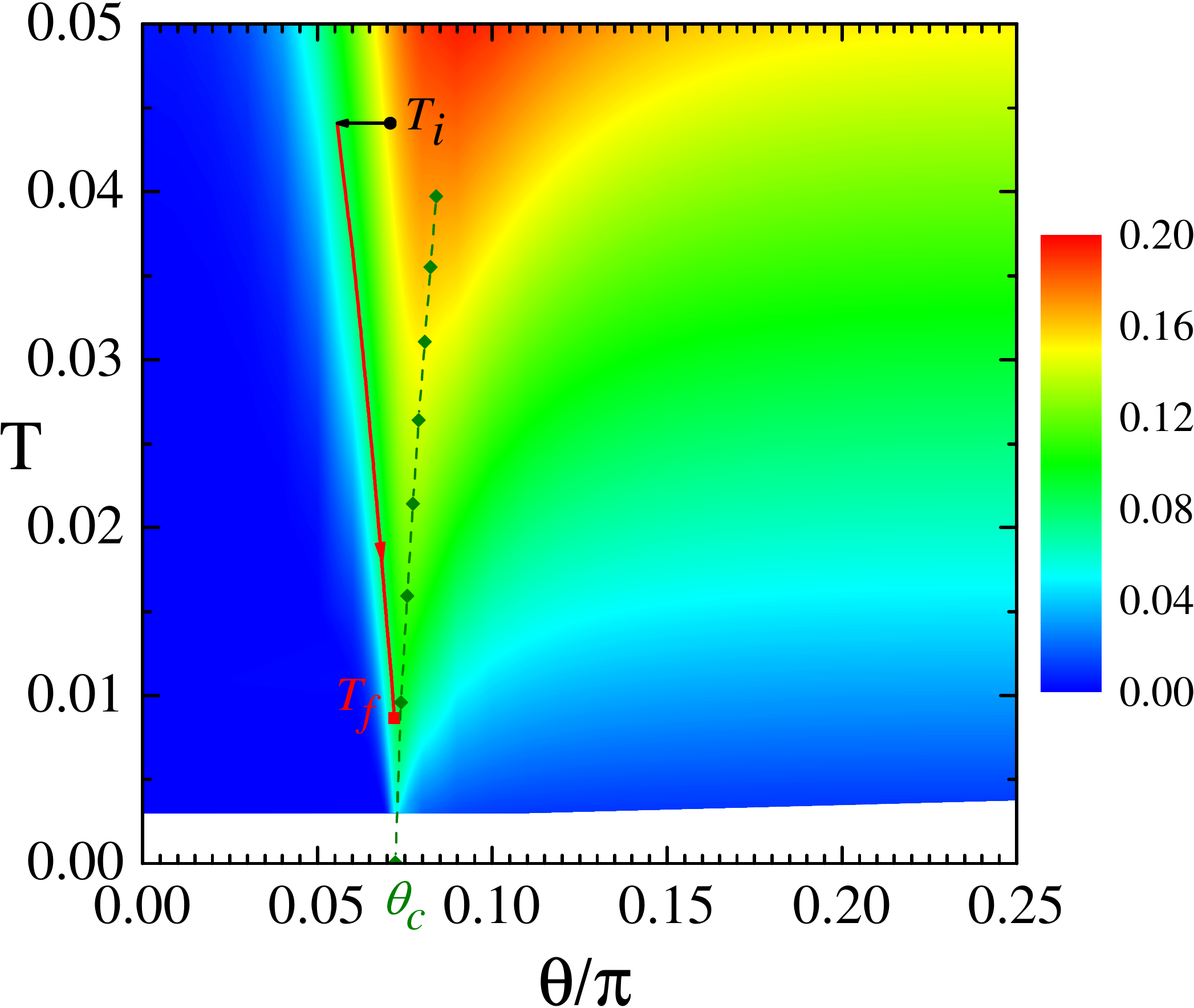}
\caption{(color online). Isentropic map of $S(T, \theta)$ at $U^\prime/U=1.2$.
The color represents the value of $S$. The transition point $\theta_c$ can be directly obtained through 
the tips ({\green $\vardiamondsuit$}) of isentropic curves. Dashed line connecting these tips is a guide to the eye.
A cooling process from $T_i$ to $T_f$ similar to MCE is shown.
The black arrow is an isothermal line with entropy decreasing, and the red arrow is an isentropic line with gap closing,
illustrating a MCE-like process.}
\label{fig2}
\end{figure}

The location of the transition point at zero temperature can be determined through the isentropic map.
It is known that, at the same temperature, the entropy $S$ has a maximum at the transition point.
As a result, all the isentropic curves should bend to the transition point.
As shown in Fig.\ \ref{fig2}, one can easily figure out that at $U^\prime/U=1.2$ the transition point locates at $\theta/\pi=0.072(1)$,
which agrees well with previous results obtained by DMRG\cite{Zhao14B}.
Meanwhile, we notice that the isentropic map shows a clear cooling process similar to magnetocaloric effect (MCE) in magnetic materials\cite{Gschneidner12}.
Staring from $T_i$ and decreasing $\theta$, one can design an isothermal process of entropy decreasing (black arrow).
Then following an isentropic curve with increasing $\theta$ (red arrow), it is allowed to decrease the temperature gradually to $T_f$.
Here the strength of SOC is used instead of magnetic field in usual MCE.
The entropy is transfered from gapless states to gapped states in the isothermal process, then followed by a gap closing in the isentropic process.
As a contrast, common MCE in magnetic materials contains an isothermal suppression of the entropy from disordered phases to FM ordered phases,
then followed by an adiabatic demagnetization \cite{Gschneidner12}.
In ultracold systems, a common way for lowering the temperature of the quantum gas
is transferring the entropy from the ground band to higher bands and removed\cite{Bark11}.
The possibility and efficiency of using the MCE-like process as an alternative technology for refrigeration in cold atom systems
need further experimental investigations.

\subsection{Specific heat}
\begin{figure} [ht!]
\centering
\includegraphics[width=.9\linewidth]{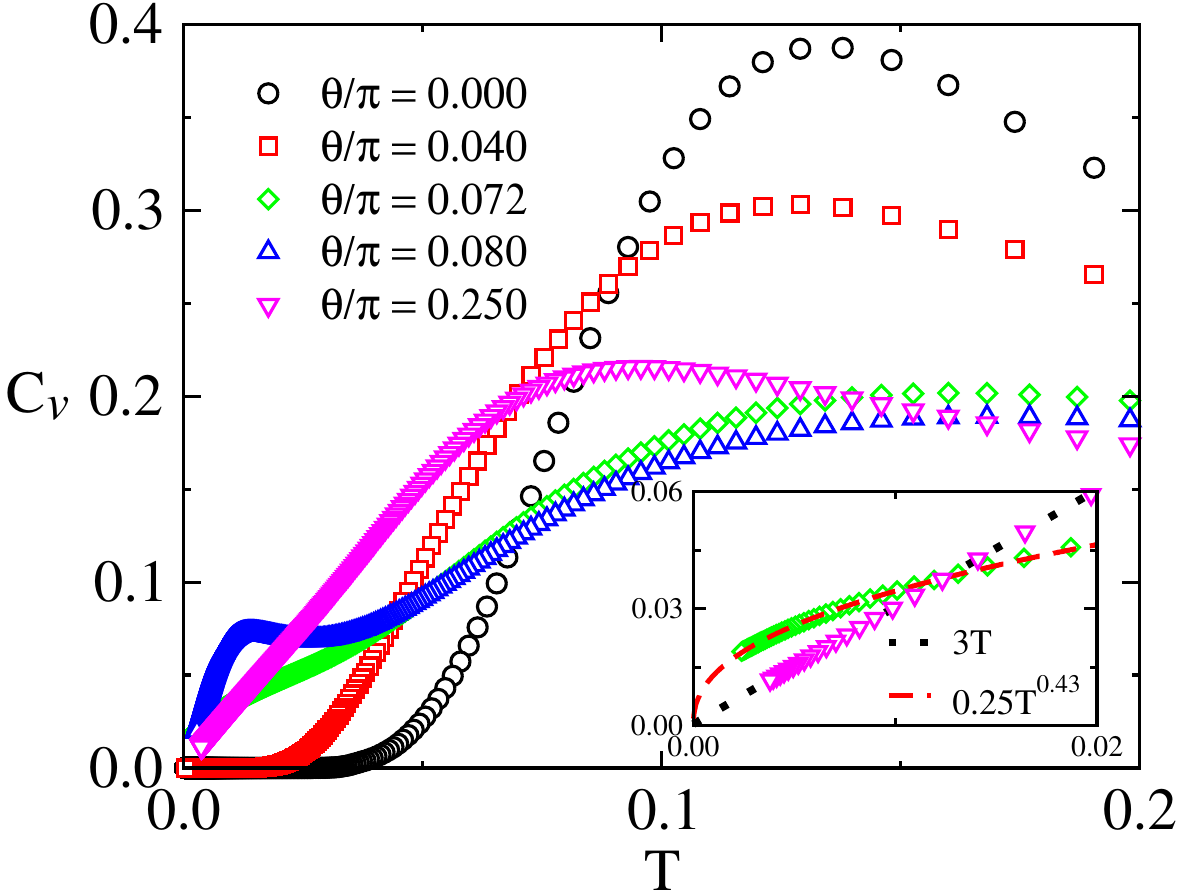}
\caption{(color online). Specific heat as a function of temperature obtained by
TMRG for a variety of $\theta$ at $U^\prime/U=1.2$. Inset: asymptotic behaviors of the specific heat at low temperature
for $\theta/\pi=0.072$ and 0.25.}
\label{fig3}
\end{figure}

In Fig.\ \ref{fig3}, we plot the specific heat as a function of temperature for a variety of $\theta$ at $U^\prime/U=1.2$,
which is qualitatively similar to the results in the exactly solvable limit.
However, at $\theta=0$, the system is now a gapped XXZ model, and the specific heat at low temperature is
$C_\nu\sim T^{-3/2}\exp(-\Delta/T)$.
At the transition point $\theta=\theta_c\simeq 0.072$, our numerical data show that it deviates from the square-root behavior,
which suggests that the dispersion of the low-energy excitations is not well approximated by $k^2$ for a finite $U^\prime/U$.
For $\theta > \theta_c$, $C_\nu\sim T$, which is a characteristic feature of TLL.

\begin{figure}[ht!]
\centering
\includegraphics[width=.9\linewidth]{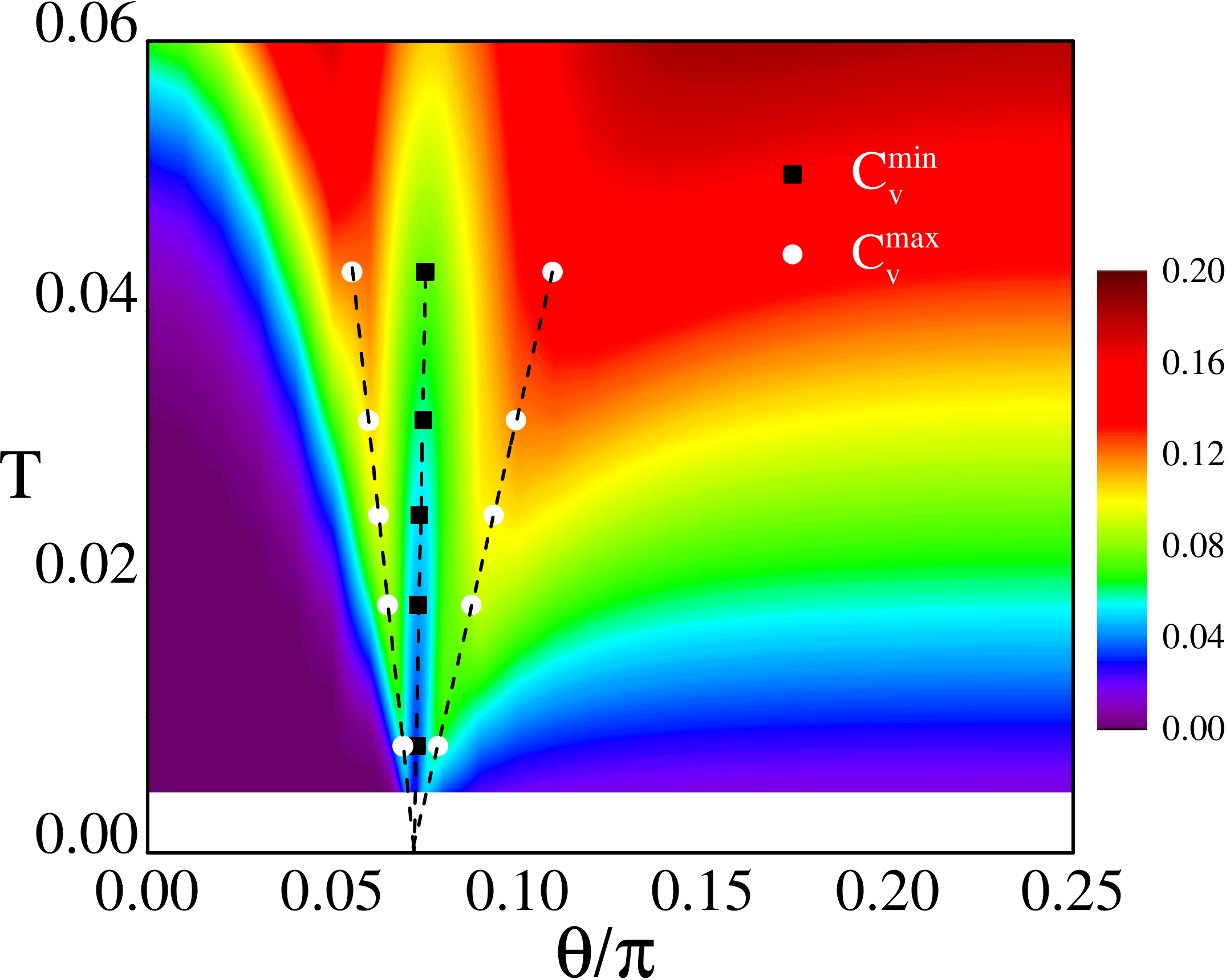}
\caption{(color online). Contour plot of specific heat $C_\nu(T,\theta)$.
Symbols are the extreme points for the given temperature and dashed lines are fitting curves.
The color represents the value of the specfic heat at given $T$, $\theta$.}
\label{fig4}
\end{figure}

The transition points can be determined from the finite-temperature scaling of the specific heat as well\cite{Vojta03,Sachdev11}.
In Fig.\ \ref{fig4}, we present a contour plot of the specific heat with $T$ and $\theta$.
For a fixed $T$, one can obtain two maxima $C_\nu^{\rm{max}}$ and  one minimum $C_\nu^{\rm{min}}$.
At these extrema, the corresponding $T_{\rm{max(min)}}$ and $\theta_{\rm{max(min)}}$ should
follow a scaling behavior\cite{Vojta03,Sachdev11}:
$$T_{\rm{max(min)}}\propto |\theta_{\rm{max(min)}}-\theta_c|^{\alpha},$$
with $\theta_c$ the transition point and $\alpha$ the critical exponent.
The fitting dashed lines in Fig.\ \ref{fig4} show rather good linear behaviors,
indicating $\alpha=1$. Furthermore, the transition point is fitted as $\theta_c=0.072(1)$,
which agrees well with that we obtain from the entropy.

Fisher and Berker have established the scaling description of classical first-order phase transitions\cite{Fisher82}.
Subsequent works show its validity in the quantum ones\cite{Kirkpatrick15B,Kirkpatrick15L}.
We notice the scaling relation $\alpha=1$ has also been found at the first-order transition point separating 
the FM phase and the TLL phase in a 1D spin-$1/2$ XXZ chain\cite{Suzuki15}.
Since the symmetry of these two model is quite different, this resemblance deserves further theoretical analysis.

\section{$\mathbf{U^\prime/U < 1}$}
As shown in the ground-state phase diagram in previous works\cite{Zhao14A, Piraud14, Xu14},
there are two phases in this case, a paramagnetic phase and a FM one.
The former is gapless while the latter is gapful. The transition between these two phases are of
BKT type\cite{KBT}. In the BKT transition, it is a big challenge to figure out the critical point accurately.
To determine the phase boundary, the entanglement entropy of the ground state of the Hamiltonian (\ref{OriginH}) 
was calculated\cite{Zhao14A, Xu14} by DMRG.
Based on their analysis, the transition seems to occur at a finite $\theta$ for a finite $U^\prime$.
Another DMRG calculation based on the effective model (\ref{HEFF}) gives a relatively large error bar
for the critical points \cite{Piraud14}. In this section, we will study the thermodynamic properties of Hamiltonian (\ref{HEFF}), 
from which we can provide solid numerical evidence that the transitions from the paramagnetic
phase to the FM phase occur at $\theta=0$. 

\subsection{Entropy}
\begin{figure}[ht!]
\centering
\includegraphics[width=.9\linewidth]{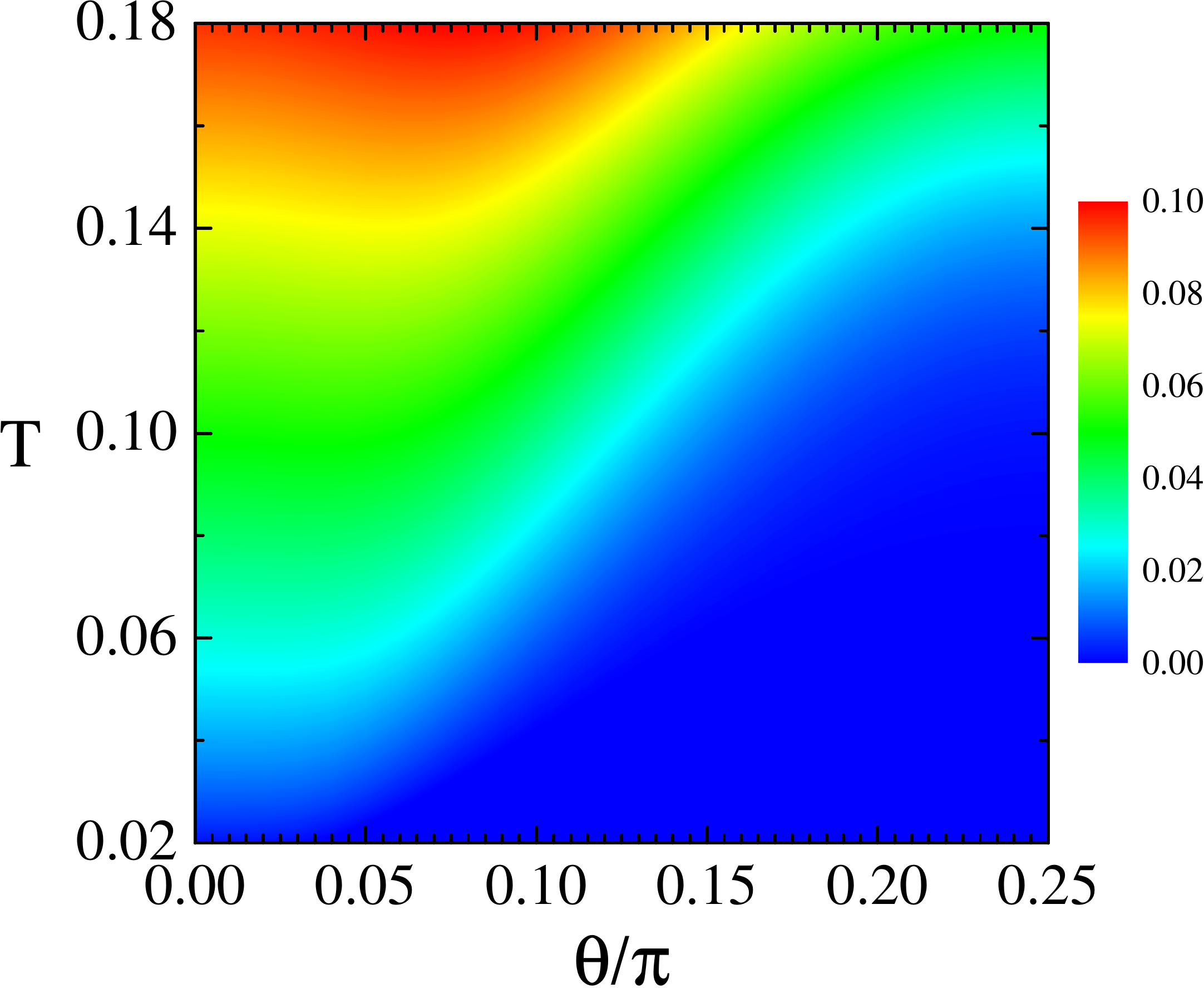}
\caption{(color online). Isentropic map of $S(T,\theta)$ at $U^\prime/U=0.5$.
The color represents the value of the entropy at given $T$, $\theta$.}
\label{fig5}
\end{figure}

In this subsection, we will discuss the entropy. For simplicity, we limit our discussion to $U'/U=0.5$.
In Fig. \ref{fig5}, we plot the isentropic map.
In contrast to the case $U'/U>1$, we do not find any singular point on the isentropic curves at finite $\theta$.
Moreover, we observe that the entropy
on the left is larger than that on the right at low temperature.
This can be understood from the known results that the FM phase is gapful while
the paramagnetic phase is gapless. As the temperature increases, the isentropic curve becomes flatter.
This is because at high temperature the thermal fluctuation dominates over the quantum fluctuation.
\begin{figure}[ht!]
\centering
\includegraphics[width=.9\linewidth]{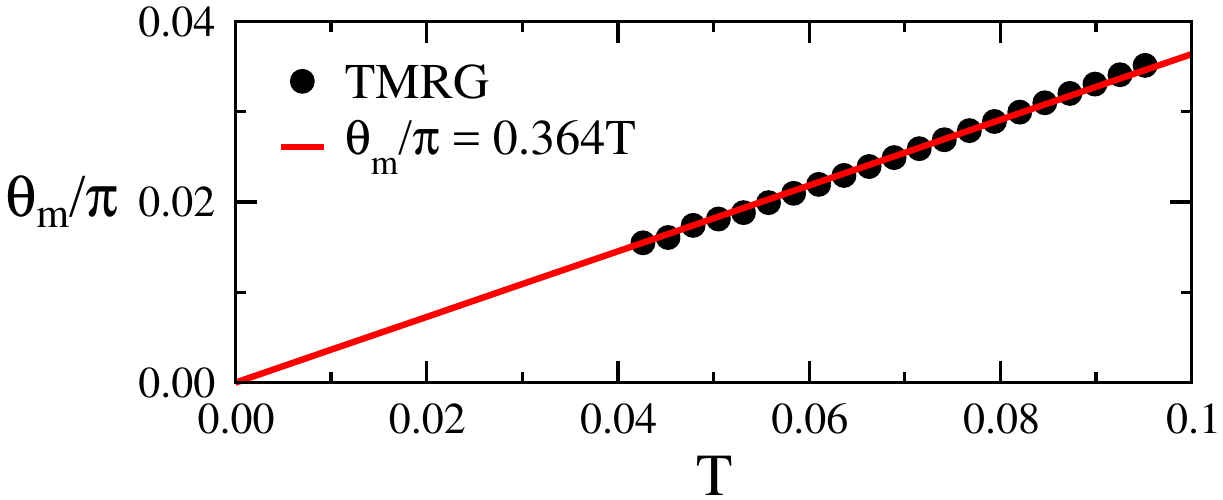}
\caption{(color online). Extrapolation of $\theta_m$ where the entropy is maximal at a given temperature to determine the critical value.}
\label{fig6}
\end{figure}
To extract the critical point between such gapped and gapless phases,
we first determine the position $\theta_m$ where the entropy is maximal for a fixed temperature,
and then extrapolate them to zero temperature.
In Fig. \ref{fig6}, we plot $\theta_m$ as a function of the temperature $T$.
The curve can be well fitted by a linear function $\theta_m/\pi = a\cdot{T}$,
with the parameter $a=0.364(2)$. Thus, we conclude that within our error bar the critical point locates at $\theta_c=0$.
In Appendix B, we perform a DMRG calculation, which confirms our conclusion further.

\subsection{Specific Heat}
\begin{figure} [ht!]
\centering
\includegraphics[width=.9\linewidth]{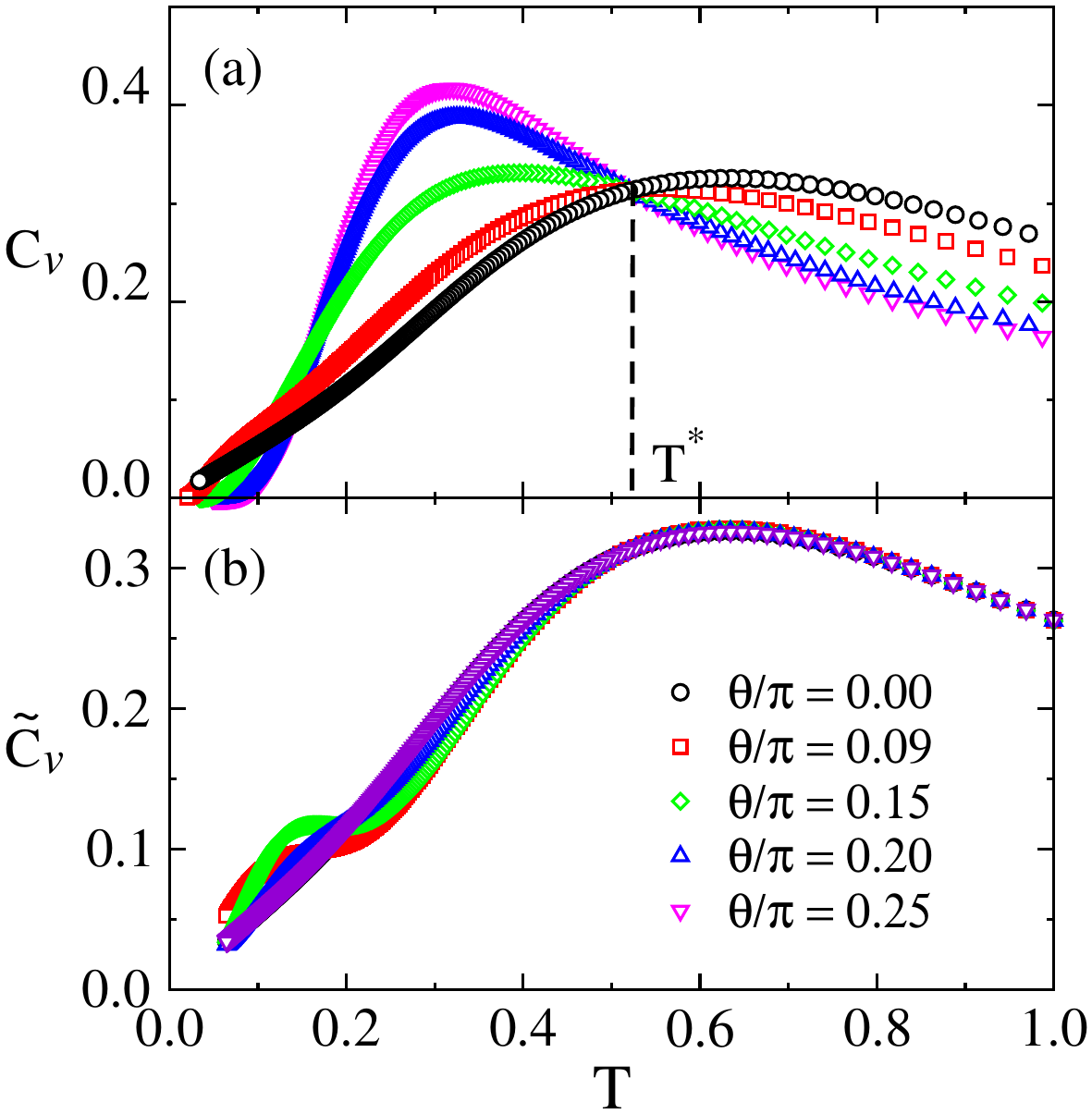}
\caption{(color online). (a) Specific heat as a function of temperature for  a variety of $\theta$ at $U^\prime/U=0.5$. 
An isosbestic point is indicated by dashed line, and the corresponding temperature is marked as $\rm{T}^*$.
(b) Scaled specific heat $\tilde{C}_\nu(T)$ collapses into one line at high temperature.}
\label{fig7}
\end{figure}

In Fig.\ \ref{fig7}(a), we plot the specific heat as a function of temperature for a variety of $\theta$ at $U^\prime/U=0.5$,
which is much different from the results of $U^\prime/U>1$.
At low temperature, $C_\nu$ decreases exponentially (linearly) in the gapped (gapless) phases.
From the exponential behavior, one can see that the energy gap increases as $\theta$ grows.
Furthermore, it is interesting to find that all the specific heat curves $C_\nu(T,\theta)$
intersect approximately at one point $T^* \approx 0.527(3)$.
Such a crossing point is called isosbestic point, which has been theoretically analyzed with $C_\nu(T,U)$ curves of Hubbard models\cite{Vollhardt97}.
This unique feature has been widely observed in many experiments, such as:
specific heat of normal-fluid $^3$He \cite{Greywall83} and heavy-fermion systems\cite{Brodale86,Adams12},
dielectric constant and optical conductivity in High-$\mathrm{T_c}$ superconductor $\mathrm{Rb_{11−x}Fe_{2−y}Se_2}$ \cite{zhe14}
and photoemission spectra of thin $\mathrm{VO_2}$ films\cite{Okazaki04}.
Following the argument given by Vollhardt\cite{Vollhardt13}, we can expand $C_\nu(T,\theta)$ as:
$$C_\nu(T, \theta)=C_\nu(T, 0)+ \cos^2(2\theta) F(T)+ O[\cos^3(2\theta)],$$
where
$$F(T)\approx \frac{C_\nu(T, \theta_1)-C_\nu(T, \theta_2)}{\cos^2(2\theta_1)-\cos^2(2\theta_2)},$$
is a function of $T$ only. The validity of this expansion can be verified by
$$\tilde{C}_\nu(T)=C_\nu(T, \theta)-\cos^2(2\theta) F(T)\approx C_\nu(T, 0).$$
As shown in Fig.\ \ref{fig7}(b), all specific heat curves for different $\theta$ collapse well into a single curve at high temperature.
We have confirmed that such isosbestic point can be observed for $U^\prime/U\gtrsim 0.45$ in our model.

\section{CONCLUSIONS}
\begin{figure} [ht!]
\centering
\includegraphics[width=.9\linewidth,clip]{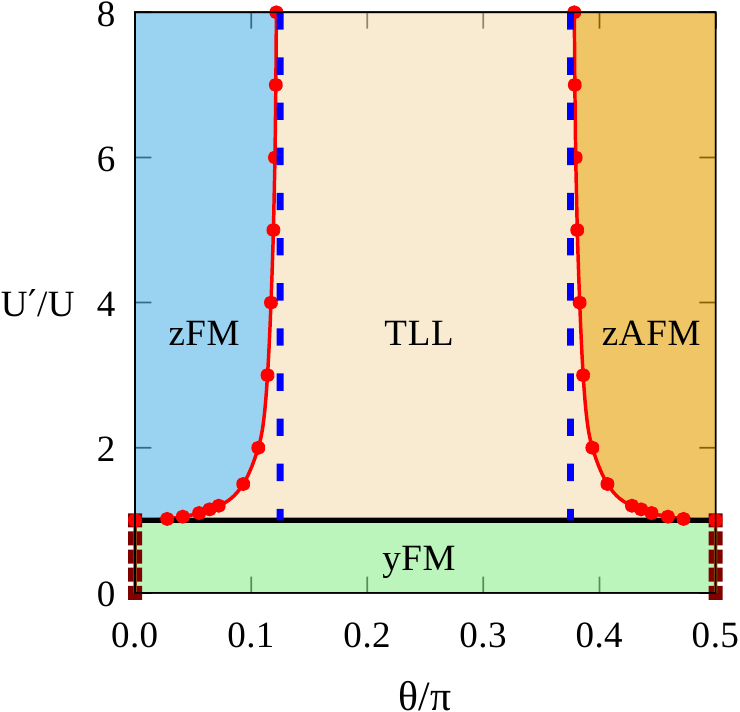}
\caption{(color online). Ground-state phase diagram of Hamiltonian (\ref{HEFF})
in the $U^\prime/U$ vs $\theta$ plane. The reflection symmetry of the phase diagram 
with respect to $\theta=\pi/4$ is the direct consequence of the transformation\cite{Zhao14A} 
for interchanging $t$ and $\lambda$.
The two exact transition points at $\theta=\pi/8$ and $\theta=3\pi/8$ in the
$U^\prime/U\rightarrow \infty$ limit are indicated by blue dashed lines. 
For $U^\prime/U>1$, the red circles are the transition points between zFM (zAF) phase and 
TLL phase, and the solid red lines are guide to eyes.
At $U^\prime/U=1$, it is a spiral phase. 
For $U^\prime/U<1$, the transition between the paramagnetic phase and the yFM phase occurs at $\theta=0$ and $\theta=0.5\pi$, 
as marked by maroon squares. The lowercase letters y and z represent the polarization directions.
}
\label{fig8}
\end{figure}

In conclusion, we study the thermodynamic properties of a spin-1/2 XYZ Heisenberg chain
with a DM interaction by using the TMRG method. This model approximates a two-component bosonic 
system with a synthetic SOC in deep insulating region. At low temperature, the asymptotic behaviors of the specific heat 
and the entropy are in close association with the properties of the ground states. 
We can thus figure out the phase boundary of the ground-state phase diagram through the isentropic map.
For $U^{\prime}/U>1$, the transition from the gapless TLL phase to the gapped FM(AF) phase occurs at a finite $\theta$.
A MCE-like process is proposed and the scaling behavior near the transition point is discussed.
On the other hand, for $U^{\prime}/U<1$, we find no sigularity in the isentropic map at finite $\theta$. 
After a careful extrapolation, we determine that the transition between the paramagnetic 
phase and the FM phase occurs at $\theta=0$ (or equivalently $\theta=\pi/2$).
We confirm this conclusion by DMRG calculations. Based on our results, a refined ground-state phase diagram is given in Fig. \ref{fig8}.

\section{Acknowledgements}
We thank Wei Li for helpful discussion.
This work was supported by the National Natural Science
Foundation of China (Grants No. 11474029, 11574200), by National Program on Key Research Project 2016YFA0300500 (X.Q.Wang), 
by the Special Program for Applied Research on Super Computation of the NSFC-GD Joint Fund, 
and by the SFB Transregio 49 of the Deutsche Forschungsgemeinschaft (DFG)
and the Allianz f\"{u}r Hochleistungsrechnen Rheinland-Pfalz (AHRP).

\appendix

\section{Exact solution in the $U^\prime/U\rightarrow\infty$ limit}
The effective Hamiltonian (\ref{HEFF}) in the limit $U^{\prime}/U\rightarrow \infty$ can be reduced to
\begin{equation}\label{effHam}
\hat{\mathcal{H}} = -J\sum_{i}\hat{S}_i^z \hat{S}_{i+1}^z - D\sum_{i}\big(\hat{S}_i^z \hat{S}_{i+1}^x - \hat{S}_i^x \hat{S}_{i+1}^z\big),
\end{equation}
with $J=2\cos\theta\geq 0$ and $D=\sin2\theta\geq0$.
The exact solution\cite{DERZHKO} of \eqref{effHam} is obtained by the Jordan-Wigner transformation.

The Hamiltonian is invariant under the rotation
\begin{align}\label{rotationHam}
    \left\{
    \begin{array}{l}
        \hat{S}^x \rightarrow \hat{S}^y \\
        \hat{S}^y \rightarrow \hat{S}^z \\
        \hat{S}^z \rightarrow \hat{S}^x
    \end{array}
    \right\}
\end{align}
and accordingly, \eqref{effHam} turns into
\begin{equation}\label{effHamRot}
\hat{\mathcal{H}} = -J\sum_{i}\hat{S}_i^x \hat{S}_{i+1}^x - D\sum_{i}\big(\hat{S}_i^x \hat{S}_{i+1}^y - \hat{S}_i^y \hat{S}_{i+1}^x\big).
\end{equation}
Using the definition
\begin{align}\label{JWtrans}
    \left\{
    \begin{array}{l}
        \hat{f}_j^{\dagger} = \displaystyle e^{-i\pi\sum_{n<j}\hat{S}_n^+\hat{S}_n^-}\hat{S}_j^+   \\
        \hat{f}_j           = \displaystyle  e^{i\pi\sum_{n<j}\hat{S}_n^+\hat{S}_n^-}\hat{S}_j^-
    \end{array}
    \right.
\end{align}
the Hamiltonian \eqref{effHamRot} finally becomes
\begin{equation}\label{effHamRot-JWtrans}
\hat{\mathcal{H}} = -\sum_{j}\left(J^{0}\hat{f}_i^{\dagger}\hat{f}_{i+1}^{\dagger} -J^{0} \hat{f}_i\hat{f}_{i+1} + J^{+}\hat{f}_i^{\dagger}\hat{f}_{i+1} - J^{-}\hat{f}_i\hat{f}_{i+1}^{\dagger}\right),
\end{equation}
with $J^{0}=J/4$, $J^{\pm}=(\frac{J}{2} \pm iD)/2$. 
After Fourier transformation
\begin{equation}\label{FourierTrans}
    \left\{
    \begin{array}{l}
        \hat{f}_k = \displaystyle\frac{1}{\sqrt{N}}\sum_je^{ikj}\hat{f}_j\\
        \hat{f}_k^{\dagger} = \displaystyle\frac{1}{\sqrt{N}}\sum_je^{-ikj}\hat{f}_j^{\dagger},\\
    \end{array}
    \right.
\end{equation} we obtain the Hamiltonian in the momentum space
\begin{equation}
\hat{\mathcal{H}}= \displaystyle -\sum_k \left[ A(k) \hat{f}_k^{\dagger}\hat{f}_k - B(k) \left(\hat{f}_k^{\dagger} \hat{f}_{-k}^{\dagger} + \hat{f}_{k} \hat{f}_{-k}\right) \right],
\label{effHamRot-JWtrans-Fourier}
\end{equation}
with $A(k)=J\cos k/2+D\sin k$, $B(k)=iJ\sin k/4$. 
The diagonalization is finished up by the Bogoliubov transformation:

\begin{align}\label{BogoliuboT}
    \left\{
    \begin{array}{l}
        \hat{\Lambda}_k = iu_k \hat{f}_k + v_k \hat{f}_{-k}^{\dagger}  \\
        \hat{\Lambda}_k^{\dagger} = -iu_k \hat{f}_k^{\dagger} +v_k \hat{f}_{-k},
    \end{array}
    \right.
\end{align}
where $u_k$ and $v_k$ are real coefficients, which fulfill the following relations
\begin{equation}\label{UkVk}
u_{-k} = -u_k, \;\;v_{-k} = v_k,  \;\; u_k^2+v_k^2=1.
\end{equation}
The transformed Hamiltonian would only contain terms proportional to $\hat{\Lambda}^\dagger_k \hat{\Lambda}_k$ when
\begin{eqnarray}\label{BogoliuBoConstrain}
\frac{J\cos{k}}{2}-\frac{J\sin k}{4}\left( \frac{u_k}{v_k} + \frac{v_{-k}}{u_{-k}}\right) & = & 0.
\end{eqnarray}
In combination with (\ref{UkVk}), we then have
\begin{equation}\label{ukvkform}
  u_k=\frac{\sin k}{\sqrt{2(1-\cos k)}},   \;\; v_k=\sqrt{(1-\cos k)/2}.
\end{equation}
Finally we end up with
\begin{equation}
\hat{\mathcal{H}}=\sum_k E_k(\hat{\Lambda}^\dagger_k \hat{\Lambda}_k-1/2),
\end{equation}
with $E_k=\cos2\theta -\sin2\theta\sin k$.

\vspace{3mm}
\section{Determining the critical points by DMRG}
To confirm our conclusion that the critical point locates at $\theta_c=0$ for $U'/U<1$,
we repeat the same calculations by Zhao et.al. \cite{Zhao14A} but for Hamiltonian (\ref{HEFF}).
The freedom at each site now is two, much smaller than that in the Hamiltonian (\ref{OriginH}), thus
allowing us to obtain more accurate numerical data as well as larger sizes.
In our DMRG calculations, we impose open boundary conditions.
500 $\sim$ 1200 states are kept to ensure the truncation errors are smaller than $10^{-7}$.
Moreover, we perform sweeps to improve the accuracy and to ensure the convergence of the ground-state energy per site 
to seven digits.

\begin{figure}[ht!]
\centering
\includegraphics[width=.9\linewidth]{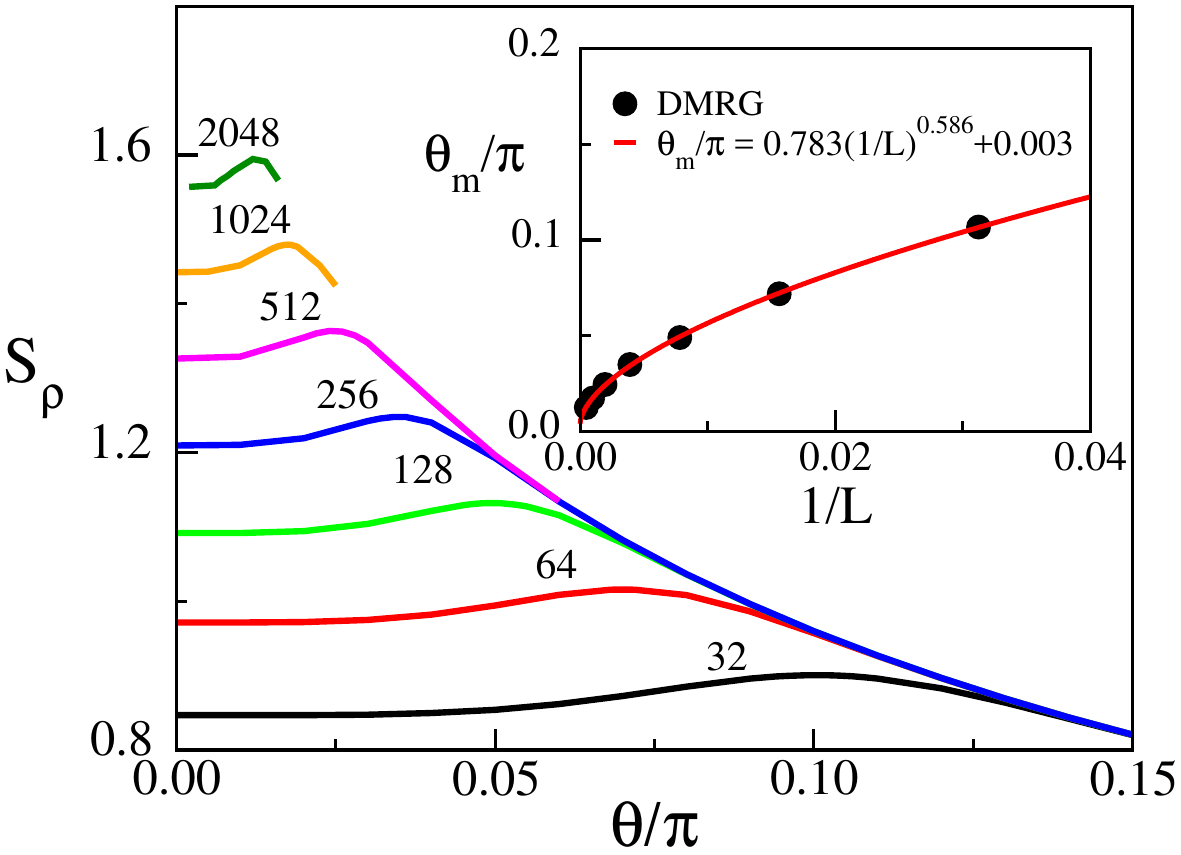}
\caption{(color online). Entanglement entropy $S_\rho$ as a function of $\theta$ for various system sizes are shown.
$\theta_m$, where $S_\rho$ shows up its maximum, are then extracted.
Inset: finite-size extrapolation of $\theta_m$ to determine the critical point $\theta_c$.}
\label{fig9}
\end{figure}

The critical point $\theta_c$ then is determined\cite{Amico08, Ejima12, Pino12} through the entanglement entropy $S_\rho$,
with $S_\rho=-\mathrm{Tr}\rho\ln\rho$ via the reduced density matrix $\rho$ of a half chain.
In Fig.\ \ref{fig9}, we first plot the entanglement entropy versus $\theta/\pi$ obtained
with various chain lengths, $L$ = 32, 64, 128, 256, 512, 1024 and 2048.
Then, we determine $\theta_m$, where $S_\rho$ is maximal, for the given length. These $\theta_m$ are extrapolated to
the thermodynamic limit with respect to $1/L$ and deduce the critical point $\theta_c$. 
In the inset, we show such an extrapolation for $\theta_m$ with a variety of chain lengths,
which can be fitted by a power-law function $\theta_m/\pi=a(1/L)^b+c$, with the best fitting 
parameters $a=0.783(4)$, $b=0.586(5)$ and $c=0.003(2)$.
Therefore, we conclude that within our error bar $\theta_c=0$. One can see that $\theta_c$ obtained by our
two different methods are well consistent.

\vfill
\end{document}